\documentclass{sigchi}

\doi{}
\isbn{}

\usepackage{balance}       
\usepackage{graphics}      
\usepackage[T1]{fontenc}   
\usepackage{txfonts}
\usepackage{mathptmx}
\usepackage[pdflang={en-US},pdftex]{hyperref}
\usepackage{color}
\usepackage{textcomp}
\usepackage{subcaption}
\usepackage{multirow}
\usepackage{booktabs}
\usepackage{array}

\usepackage{microtype}        

\usepackage{todonotes}

\def\plaintitle{Using machine learning to build public policy agenda from social media conversations}

\def\emptyauthor{}
\def\plainkeywords{Machine learning; public policy; agenda setting; social media data.}

\makeatletter
\def\url@leostyle{%
  \@ifundefined{selectfont}{
    \def\UrlFont{\sf}
  }{
    \def\UrlFont{\small\bf\ttfamily}
  }}
\makeatother
\def\pprw{8.5in}
\def\pprh{11in}

\definecolor{linkColor}{RGB}{6,125,233}
\hypersetup{%
  pdftitle={\plaintitle},
  pdfauthor={\emptyauthor},
  pdfkeywords={\plainkeywords},
  pdfdisplaydoctitle=true, 
  bookmarksnumbered,
  pdfstartview={FitH},
  colorlinks,
  citecolor=black,
  filecolor=black,
  linkcolor=black,
  urlcolor=linkColor,
  breaklinks=true,
  hypertexnames=false
}

\makeatletter
\def\@copyrightspace{\relax}
\makeatother

\begin{document}

\title{\plaintitle}

\numberofauthors{3}
\author{%
  \alignauthor{Rahman Sanya\\
    \affaddr{Makerere University}\\
    \affaddr{P. O. Box 7062}\\
    \affaddr{Kampala, Uganda}\\
    \email{rahman.sanya@mak.ac.ug}}\\
}

\maketitle

\begin{abstract}
 
Issue identification and agenda setting represents an important stage in the public policy making process. Traditional approaches for carrying out activities under this stage are time- and labor-intensive on data collection and analysis, in addition to being costly to scale over large geographic areas. In this work we propose a human-augmented machine learning (ML) approach for identifying matters of public interest from social media conversations. The approach consists of five stages namely, input data cleaning and pre-processing, keywords extraction and issue identification, narrative creation, narrative validation, and agenda validation. We implemented experiments to validate the output of our method based on a Twitter dataset and using Latent Dirichlet Allocation (LDA) and Top2Vec for topic modeling. Natural Language Generation (NLG) was achieved using GPT-2 while narrative and agenda validation were based on similarity analysis and human evaluation. We achieved "very good" and "good" inter-rater agreement (IRA) on readability and coherence of agenda narrative generated by our GPT-2 model. On the other hand, IRA was "good" for generated agenda items. We also achieved above average cosine similarity score on at least three out of five reference text (narrative) themes. These results demonstrate that the ML approach represents a promising methodology for identifying issues of public interest from social media conversations.
\end{abstract}


\section{Introduction}

Issue identification and agenda setting is the first, and one of the most important phases in the public policy making process. During this phase, important social problems affecting society are identified and considered for placement onto the public policy agenda. A public policy may be defined as the set of plans, courses of action, and decisions a government has decided to take to solve or deal with a problem, while an agenda represents a temporally organized list of matters to be attended to. What are the most important social problems affecting society today, and which of them gets onto the public policy agenda is considered a kind of competition and a contest for limited government attention and resources. It is therefore, important that only legitimate issues that serve the general public interest are identified and get onto the public policy agenda. An important question in this regard is, how do we go about identifying issues that truly serve the public interest for inclusion on the policy agenda?

Current literature identifies three main ways by which issues make their way onto the public policy agenda. Here, issue broadly refers to a real-world situation that falls short of public expectation. A matter (or issue) is likely to get onto the public policy agenda if it is pushed there by an influential constituent or group, when an unexpected event occurs (e.g., a pandemic), or due to unusually high public interest and outcry. In the latter case, the news media often plays an important role in amplifying the issues and hence, pushing them onto the public policy agenda \cite{Dearing1997}. In recent times however, public interest advocacy groups have taken on an increasing role of driving matters onto the policy agenda. For example in Uganda, civil society organizations were instrumental in mobilizing and gathering views from grassroots people and publishing them in what they called the Citizens Manifesto 2016-2021 \cite{UGMP2015}. This document is a compilation of issues considered pertinent by members of the public in Uganda prior to the 2016 general elections. The process used to gather views from stakeholders involved carrying out opinion surveys, which was achieved by holding countrywide workshops. The collected data was then analyzed manually using qualitative data analysis methods to identify salient issues before these were validated through further consultations at national level. This approach has several advantages including the fact that surveys can be designed objectively so as to collect representative data sets across national geographies and demographics (thus eliminating data bias). It is also a comprehensive, actively participatory, and inclusive process. However, this kind of approach may be laborious and/or costly to scale-up over large geographic areas, and is inefficient on both data collection and analysis activities. Results from such opinion surveys may also become obsolete as soon as they are published since it often takes a long time to complete the data collection, analysis, validation, and reporting cycle. This may thus necessitate a new survey to obtain up-to-date insight on social problems needing public policy attention. Furthermore, such traditional approaches require extensive human expertise to be successful, for example trained data collectors and analysts. These kinds of expertise may be scarce in some low-income countries.

Social networking platforms such as Twitter and Facebook are increasingly becoming important global spaces for citizens to not only mobilize each other for public causes but also provide a vital tool for voicing opinions on matters of public interest. For example, some studies have shown that there is a positive relationship between online and offline political activity \cite{Ahmad2019, Vaccari2018}. Others have argued that these platforms offer alternative spaces for political participation \cite{Waller2013}, for gauging political support \cite{Fang2019}, analyzing discussion themes \cite{Fang2019, Jordan2018}, and predicting engagement in political talk \cite{Shugars2019}. Apart from the preceding use cases, social media data have also been used for research in other domains including public health e.g., for pandemic surveillance \cite{Krupenkin2019, Burzynska2020}, communication and education in healthcare \cite{Wong2020}, infodemic studies \cite{Cinelli2020}, and pandemic policy decision-making \cite{Yigitcanlar2020}. A literature review of social media data use in this domain can be found here \cite{Tsao2021}. In context of the current work, social media has been used to influence public policy. For example Hunt \cite{Hunt2020} investigated the use of Twitter to influence public policy debate with respect to ultra-processed foods.

A variety of methods may be used to analyze and derive insight from social media data. In recent times however, machine learning (as opposed to, say, data mining) approaches have dominated this space. For example, Stukal et al \cite{Stukal2019} developed a Deep Neural Network (DNN) classifier consisting of an ensemble of four component classifiers (support vector machines, ridge regression, extreme gradient boosting tree, and AdaBoost) to determine the political orientation of bots on Twitter. Topic modeling was used by Sterling et al \cite{Sterling2019} to investigate similarity and dissimilarity in political ideologies. Ceron et al \cite{Ceron2019} used integrated sentiment analysis techniques to determine support for and against a religious ideology while Fang et al \cite{Fang2019} employed Naive Bayes to understand preference for political candidates.

From our survey of the literature, no studies have attempted to investigate the suitability of machine learning (ML) methods and social media data for the purpose of advancing understanding of public policy agenda setting. This work aims to partly address this knowledge gap, especially in the context of low-income countries in Sub-Sahara Africa. The research question we sought to answer is thus: how can we apply ML methods to social media data for purpose of identifying pertinent issues and for building a public policy agenda?

The goal of our work is to employ ML in conjunction with social media data for identifying relevant issues for inclusion onto public policy agenda. Specifically, we used topic modeling to uncover important themes in a Twitter dataset. Topic Modeling (TM) uses unsupervised ML methods for discovering latent semantic structure i.e., extracting meaningful themes from large collections of text data. Based on the identified themes, we then used a text generation method (i.e., Generative Pre-trained Transformer 2 or GPT-2) to create agenda item narratives. The third and fourth stages in this process involved validation of narrative texts and the agenda items by relying on human expertise, respectively. The contributions of our work include the following;

\begin{enumerate}
	\item We propose an approach based on ML for building public policy agenda from social media conversations. A major benefit of the proposed approach is that it allows for efficient generation and processing of large quantities of policy-related input data from a wide range of sources from the Web. This work thus demonstrates that ML methods can be used to compliment and/or validate public policy agenda and prioritization derived by alternative methods.
	
	\item An end-to-end ML pipeline is proposed for semi-automated (i.e., human-augmented) public policy agenda building based on social media data sets. The pipeline consists of five stages namely: data cleaning and pre-processing, keywords extraction, narrative text generation, narrative text validation, and agenda validation.
	
	\item A proof-of-concept ML model has been developed to validate the output of our proposed approach in 1) above. Based on input data from Twitter and using Uganda's 2020 / 2021 general elections as case study, the model provides useful insight on social issues raised by members of the general public during the elections period.
	
\end{enumerate}

The rest of this paper is organized as follows. We present related work in Section \ref{relatedwork} and materials and methods in Section \ref{materialsMethods}. Results, discussion, and conclusion are found in Sections \ref{results}, \ref{discussion}, and \ref{conclusion}, respectively.

\section{Related Work}
\label{relatedwork}

In the recent past, ML methods in conjunction with social media, Web, or other publicly available data sources have been applied on diverse tasks to solve real-world problems. In this section we review recent literature with a focus on three things: application domain and/or task, ML method(s) used, and data sources.

Public health is one of the domains in which ML methods have been applied widely for a variety of tasks, with the dominant application area currently being disease surveillance. An early work in this area is that of Ginsberg et al \cite{Ginsberg2009}, who used Web search engine query data to detect epidemics of influenza-like illnesses. The main ML method they employed is linear regression. More recently, Kaveh-Yazdy and Zarifzadeh \cite{Kaveh-Yazdy2021} used topic modeling on Telegram data to track Iran's national COVID-19 response committee's major concerns. An exploratory analysis of COVID-19 tweets using a variety of topic modeling techniques is provided by Ordun et al \cite{Ordun2020}. Deng et al \cite{Deng2013} proposed a method for analyzing the evolution of public concerns in social media. Their method reconstructs the topic space of 'expandable' micro-blogging messages, and tracks the movement of their re-posts as a form of 'public concern'. Effectiveness of the method is tested by applying it to H7N9 bird flu data collected from Weibo, a Chinese micro-blogging website. Another work in this domain that utilizes topic modeling techniques and user network analysis based on Twitter data is \cite{Pirri2020}.

ML techniques have also been applied to tasks in other domains including for political sentiment analysis, understanding user characteristics and/or interests, and tracking evolution of public concerns. For example, the work of Gasparetti \cite{Gasparetti2017} employed clustering techniques (e.g., LDA and SLA) to model user interests based on Web browsing data. Topic modeling in conjunction with Twitter data was used to characterize citizens by Vargas-Calderon and Camargo \cite{Vargas-Calderon2019}. Work on analysis of political sentiments or scheming based on Twitter data can be found in Jin et al \cite{Jin2017} and Wang et al \cite{Wang2012}. The topic modeling and sentiment analysis techniques used included TF-IDF, word2vec, doc2vec, and naive Bayes. A review of approaches for topic detection from Twitter data is provided by Mottaghinia et al \cite{Mottaghinia2020}.

Our review of the literature has thus shown that a wide range of topic modeling techniques exist, and have found applications in diverse domains. However, to the best of our knowledge no study has explored their effectiveness for the task of public policy agenda building from social media conversations.

\section{Materials and Methods}
\label{materialsMethods}

\subsection{Materials}

\subsubsection{Study area and context}
\label{studycontext}

The study area for this work is Uganda, a low-income country in Sub-Sahara Africa. Uganda had a projected population of 41.5 million people as of year 2020 \cite{UBoS2016}. According to the website www.datareportal.com \cite{Kemp2020} there were 2.5 million social media users in Uganda by January 2020, putting penetration at 5.6\%. Facebook (41.55\%), Twitter (24.72\%), and Pinterest (23.64\%) were the three most popular social networking sites in Uganda between January 2020 and January 2021, according to the website www.statcounter.com \cite{statCounter2020}.

The social media data set used in this work relates to Uganda's 2020/2021 general elections. During those elections, the country adopted what it called 'scientific campaigns' so as to prevent the spread of corona virus disease 2019 (COVID-19). Although not defined, scientific campaigns were generally understood to mean mobilizing and addressing small crowds of up to 200 persons while ensuring observance of COVID-19 prevention measures including social distancing, hand hygiene, and wearing a facial cloth mask. Furthermore, Uganda's elections body encouraged candidates vying for political positions to make use of Information and Communication Technologies (ICT) to carry out their campaigns. Among others, candidates were encouraged to use social media platforms like Facebook, Twitter, and YouTube to promote their political agendas and to reach out to voters. In addition to responding/reacting to, or promoting the campaign messages of their favorite candidates, social media users also used the same platforms to disseminate messages about issues they considered important. For example, \#StaySafeUG, \#WearAMask, and \#COVID19 were some of the hashtags that trended in Uganda among Twitter users throughout the general elections period (see sub-section \ref{dataset}). These hashtags were reminders urging the Ugandan population to keep observing health guidelines to protect themselves from COVID-19. We therefore, considered the context and timing (i.e., general elections) of our social media data collection activities appropriate for this public policy related study.

\subsubsection{Social media data collection}

Our study is based on use of a Twitter dataset, which was collected by ourselves over the period from November $2^{nd}$, 2020 to January $8^{th}$, 2021, when Uganda held its general elections. To aid the data collection exercise, we identified five neutral hashtags that were associated with these elections and campaigns. The hashtags were neutral with respect to political parties and/or candidates and included, \#UGPresNominations, \#UgandaDecides2021, \#UGDecides2021, \#UGVotes2021, and \#Uganda. Tweet data collection was scheduled on a weekly basis for a period of 10 weeks. During each week, the hashtags used for tweet extraction were reviewed and updated where necessary to include new ones that had come up and/or removed others that had been overtaken by events. For example, the \#UGPresNominations hashtag did not seem useful anymore weeks after the presidential nominations. Information extracted from each tweet included user identifier, date of user account creation, user screen name or handle (i.e., the @name), account description, user location, following count, user follower count, tweet count, tweet timestamp, tweet identifier, tweet coordinates, tweet place, retweet count, favorite count, hash tags, and tweet text. A total of 56,737 tweets were collected over the 10 weeks period. We used Tweepy (www.tweepy.org/), an open source and free Python library to scrape tweets based on the standard Twitter API.

\subsection{Methods}

\subsubsection{Exploratory analysis of our Twitter data set}

We performed exploratory analysis of our data set to understand its characteristics prior to using it for experiments. Our strategy involved analyzing the data set for each week separately, and then analyzing the combined data from all weeks. Analyzing the data by week ensured we did not lose information about important events that did not occur frequently, as there was evidence of this when we compared weekly and overall analysis results. Preventing information loss during analysis was vital in identifying issues (based on keywords) relevant for building a public policy agenda. The kinds of analyses we performed included text statistics, hashtag frequency, term/word frequency, bigram/trigram analysis, and time series analysis.

Under text statistics we performed tweet character length analysis. This analysis is important as it can affect how one represents the text as features for ML models, e.g., Term Frequency-Inverse Document Frequency (TF-IDF) is usually too sparse for short texts and average Word2Vec is usually too noisy for long texts. Text length can also affect the algorithm used e.g., Long Short Term Memory (LSTM) are better than vanilla RNN networks on long texts. Term and hashtag frequency analyses were useful for determining which words and hashtags were most used in tweets, respectively. Bigram and trigram analysis helps to identify words that frequently occurred together in groups of two or three, respectively. Time series analysis was done to determine temporal topic distribution. A number of Natural Language Processing (NLP) libraries based on the Python programming language were utilized for exploratory data analysis (EDA) including Gensim and the Natural Language Toolkit (NLTK). Data visualization was done using histograms, bar graphs, network diagrams, and word clouds (used for understanding common words used in tweets).

\subsubsection{Machine learning pipeline for agenda building from social media data}

\begin{figure}[h!]
	\centering
	\includegraphics[width=0.45\textwidth]{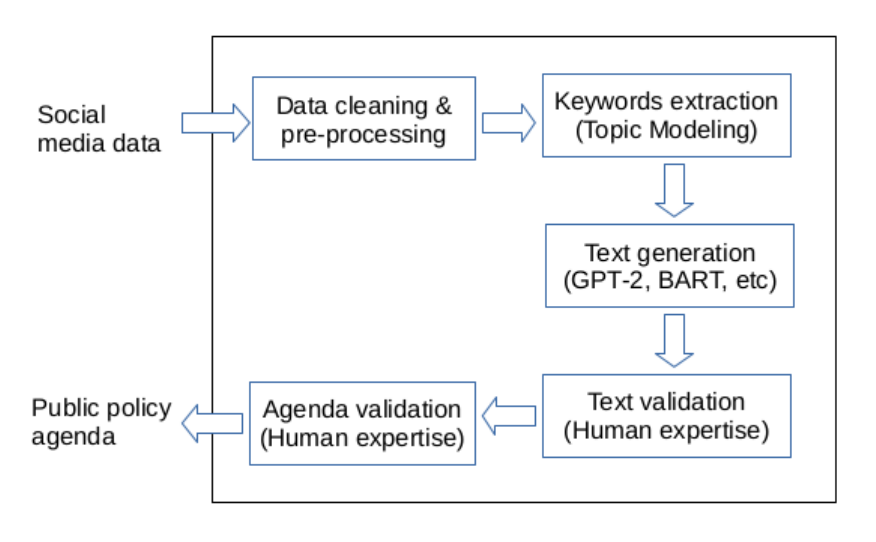}
	\caption{Human-augmented ML pipeline for generating public policy agenda from social media data.}
	\label{fig:pipeline}
\end{figure}

Cobb et al \cite{Cobb1976} define agenda building as the process by which the needs of different constituents in a society become issues competing for the attention of public officials. In this work, we propose a semi-automated (i.e., human-augmented) pipeline or workflow for agenda building based on ML and social media data. The assumption we make here is that all interest groups will utilize social media (or the Web) to voice and/or to disseminate information about issues they consider pertinent for public policy attention. The pipeline consists of five stages namely data cleaning and pre-processing, keywords extraction and issue identification, text generation (i.e., narrative creation), text (or narrative) validation, and agenda validation. The pipeline is shown in Figure \ref{fig:pipeline}. Each stage is described in the following sections.

\textbf{Data cleaning and pre-processing}

Unprocessed social media data is generally noisy in nature, consisting of many unwanted alphabetic characters and symbols, in addition to other elements such as emoticons, URLs, images, video clips, hashtag symbol (\#), user mention symbol (@), etc. A single social media message may also contain content in multiple languages. The purpose of the data cleaning and pre-processing stage is thus to clean the data set of all unwanted elements and symbols. It is also used to prepare the data set in a format suitable for language modeling. This is especially important since we used pre-trained word embeddings (i.e., pre-trained networks) for our topic modeling since these have built-in vocabularies or dictionaries. Cleaning activities  performed included removing retweets, twitter handles, URLs, punctuation marks, numbers, special characters, non-English words, and alphanumeric words. Other elements removed included collection words (i.e., search phrases) and short words (words less than three characters long). Stop words i.e., words that carry no informative value were also removed. In addition to the English stop-words defined in the Python NLTK package \cite{Loper2002}, we included other stop-words that occurred frequently in our text corpus.

Pre-processing our data set for TM involved tasks such as lower casing, tokenization, stemming (we did not use lemmatization), vectorization, and building word embeddings. The tool we used for cleaning and pre-processing the data set is the Python library NLTK.

\textbf{Keywords extraction and issue identification}

The goal of this stage is to extract general keywords from the processed input social media data and from these, identify issue-specific keywords that are relevant for public policy attention. Issue-specific keywords are words or phrases that have a semantic association with a public policy matter. For example for a public policy matter such as healthcare, relevant keywords or phrases might include health insurance, medicine, health worker, COVID-19, and vaccine. To extract the general keywords from social media data one uses TM.

While there are many TM methods available, we used Latent Dirichlet Allocation (LDA) and Top2Vec. LDA \cite{Blei2003} is one of the most widely used traditional methods for TM based on bag-of-words (BoW) representation of documents. These traditional methods require the user to have apriori knowledge of the number of topics or themes to be extracted from the text corpus. To determine an optimal number of topics for our work, we built multiple LDA models with differing number of topics and then computed their coherence scores, before selecting the model that gave meaningful and interpretable topics. A coherence value is used to quantify the coherence of a set of statements in a text \cite{Roeder2015}. Top2Vec \cite{Angelov2020} on the other hand, is a recent TM method based on joint distributed representation of words, documents, and topic vectors (word embedding) i.e., it is a representation learning method similar to word2vec \cite{Mikolov2013} and doc2vec \cite{Le2014}. Unlike the latter two methods, top2vec has the benefit of discovering topics that are more informative and representative of the data set it is trained on. Training a top2vec model requires three configuration settings including document (the input corpus), workers (number of worker threads), and speed. The latter configuration determines how fast the model trains and the quality of output vectors. We set the 'speed' parameter option to 'deep-learn' (as opposed to 'fast-learn' or 'learn') to ensure the best quality vectors were learned. Unlike the LDA method, top2vec is able to determine the number of topics automatically, does not require stop-word lists, stemming (or lemmatization), and the kind of pre-processing described in the previous section.

After extracting keywords, we identified dominant topics, their percentage contribution in each tweet, the most representative tweet for each topic, and word clouds for top $n$ keywords in each topic. To visualize the discovered topics along with their corresponding keywords we used the Python library pyLDAvis \cite{Sievert2014} and word clouds.

\textbf{Text generation (or narrative creation)}

The goal of this stage is to produce human interpretable text paragraphs (i.e., narratives or short stories) that convey meaningful sentiments about an issue of public concern. Our interest and strategy is that given a keyword or phrase, the model should be able to create grammatically correct and coherent phrases and/or statements (to appropriate detail) that are consistent with the meaning or theme of the initialization keyword or phrase. For this task, a number of text generation frameworks existed including Progressive Generation of Text (ProGeT) by Tan et al \cite{Tan2020}, Generative Pre-trained Transformer 3 (GPT-3) by Brown et al \cite{Brown2020a}, Generative Pre-trained Transformer 2 (GPT-2) by Radford et al \cite{Radford2018}, and BART by Lewis et al \cite{Lewis2019}. The latter is a sequence-to-sequence model while the rest share the same underlying network architecture. The interested reader is referred to the respective literature sources for information about the architecture and capabilities of each framework.

In this work we used GPT-2, a large-scale general purpose unsupervised language model. Predecessor to GPT-3, GPT-2 is capable of reading comprehension, machine translation, question answering, and text summarization. We chose to use GPT-2 because smaller versions of this very large model have been open-sourced, which has allowed development of Python packages for fine-tuning it on domain-specific data sets.

The authors of GPT-2 have released multiple pre-trained models namely the 124M or small model (with 124 million parameters), the 355M or medium model, the 774M or large model, etc. Our experiments were implemented using the 124M model for the reason that it is lightweight, faster, and allows for fine-tuning on custom data sets.

Our experiments were implemented using the Python programming language library gpt-2-simple found at this website https://github.com/minimaxir/gpt-2-simple, running on Google Colaboratory. The latter is a free cloud computing research platform equipped with Graphics Processing Unit (GPU) capability. To fine-tune GPT-2 on our data set, we trained it for 1,000 steps. At least five experiments were completed. To produce text from the trained model we configured it as follows: text length 50, temperature 0.7, top\_k 40, and top\_p 0.9. While the former parameter was arrived at through experimentation, the rest were default values. The text length of 50 gave us the best results since our interest was in generating short texts and/or phrases consistent with the capabilities of GPT-2. The results of this stage are reported in sub-section \ref{modelagendatext}.

\textbf{Text (or narrative) validation}

The objective of this stage is to evaluate the meaningfulness or interpretability (e.g., coherence) of the text paragraphs produced by the ML model during the previous stage. Justification for this stage is that the meaning of a statement in a given human language is dependent on many factors, including usage context, grammar (e.g., punctuation and spelling), etc. Multiple automated metrics are available for evaluating the quality of text generated by Natural Language Generation (NLG) systems. Examples of these are MS-Jaccard \cite{Montahaei2019}, Harmonic Bilingual Evaluation Understudy (BLEU) \cite{Shi2018}, and Metric for Evaluation of Translation with Explicit Ordering (METEOR) \cite{Banerjee2005, Lavie2007}. However, automated metrics have been shown to have poor correlation with human judgment \cite{Liu2016a, Novikova2017}. Human evaluators of various kinds (e.g., experts \cite{Belz2006} and crowd-sourced annotators \cite{Kryscinski2019}) have also been used as an alternative to automated metrics to assess the quality of text. A detailed survey of evaluation metrics for NLG systems is found here \cite{Sai2020}.

We employed human expertise to evaluate text generated by the GPT-2 model because it is the primary approach for evaluating NLG systems, even though it is currently considered less reliable due to inconsistent user ratings. However, it has been demonstrated \cite{Santhanam2019, Novikova2018} that the quality of human judgments can be enhanced through appropriate experiment design to address the problem of low reliability.

We recruited second and third year students taking bachelors study programs with English language as a major to evaluate text samples generated by our GPT-2 model. A Web-based 4-point Likert scale evaluation scheme was designed for this purpose. The scheme utilized two metrics: readability and coherence. Readability or fluency is a measure of linguistic quality of text \cite{Gatt2018, Novikova2017}. Coherence in our case is ability of the NLG system to produce text that has themes that are consistent with each other. The four points on our evaluation scale (i.e., non-overlapping categories of our nominal response variable) thus corresponded to: not coherent at all (0), not coherent (1), coherent (2), and very coherent (3). A similar 4-point scale was defined for readability. Randomly selected text pieces generated by our model were evaluated by non-unique raters selected randomly from a larger population. Fleiss' Kappa,  found to be comparable to Krippendorff's alpha (see \cite{Zapf2016}), was computed to separately assess inter-rater agreement for coherence and readability. The study design followed guidelines in \cite{Statistics2019}. Results of evaluating our text generator model are found in sub-section \ref{validatedmodeltext}.

\textbf{Agenda validation}

This stage aims to determine the validity of public policy agenda items derived from text paragraphs or narratives created using the ML model in the previous stage. The key question this stage tries to answer is: does information or themes contained in the pieces of text or narratives formed by the model represent valid social concerns within the context being investigated? Multiple different approaches may be considered to establish the validity of agenda items created using an ML model as is the case for the previous stage. However, we think that the use of domain expertise should be the primary approach because currently no reliable alternative methods exist for this kind of task. Two strategies for human evaluation of public policy agenda items generated by an ML model would include the use of policy analysts or, crowd-sourcing public opinion through a survey. Another approach for validating ML generated public policy agenda would involve the use of methods for analyzing text similarity, especially on the basis of semantics. Here, text for agenda items generated by the model are analyzed and compared with corresponding reference texts on the basis of semantics, meaning, and context to determine how close or far apart the two are. Similarity between the two texts is then quantified and measured by an appropriate metric. A survey of similarity measures for text documents can be found in \cite{Wang2020}.

In this work, we employed both human expertise and methods for studying document similarity to validate public policy agenda derived by our ML model. Reference texts for similarity analysis were extracted from manifesto documents of two leading political parties during Uganda's 2020/2021 general elections. Cosine similarity was used to analyze similarity between GPT-2 derived agenda texts  and the reference texts. Cosine similarity is computed as shown in Equation \ref{cosine-similarity},

\begin{equation}\label{cosine-similarity}
cosine(x,y)=\frac{x.y^T}{||x||.||y||}
\end{equation}

where $x$ and $y$ are vectors for a document in our GPT-2 generated agenda items and a corresponding reference text, respectively. The cosine similarity score value will range from 0 (lowest similarity) to 1 (highest similarity). Our reference texts, in form of keywords or phrases, were extracted from manifesto summaries provided on the Uchaguzi website at https://uchaguzi.go.ug. Uchaguzi is a Uganda government Information and Communication Technology based civic education platform.

For human assessment of our model generated agenda items, we crowd-sourced public opinion. A Web-based 4-point Likert scale survey tool was developed for this purpose, similar to the one described in the previous sub-section. Respondents were asked to rate each statement on whether it contained important themes for public policy attention in Uganda. Randomly selected text pieces generated by the model were evaluated by non-unique raters selected randomly from a larger population. We made minimal adjustments to a few of the statements by correcting for context, repetitions, grammar, spelling, and punctuation. For example, a statement like "Agriculture minister intervenes as BHC asks government to consider increasing funding for the sector" was re-written to "Agriculture: government should consider increasing funding to the sector." These corrections were necessary to help the respondents focus on meaning of the statement and not be distracted by syntactical errors. Inter-rater agreement was assessed using Fleiss' Kappa. Results for agenda validation are presented in sub-section \ref{validatedagenda}.

\section{Results}
\label{results}

\subsection{Dataset}
\label{dataset}

\begin{figure}[h!]
	\centering
	\includegraphics[width=0.5\textwidth]{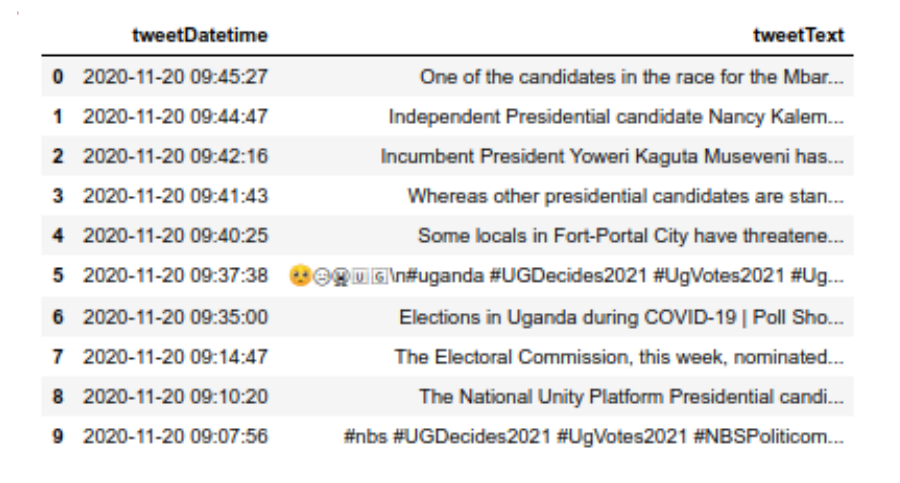}
	\caption{Example uncleaned tweets from our dataset.}
	\label{fig:sampleuncleantweets}
\end{figure}

Our final data set consists of a total 12,272 uncleaned tweets after removing retweets i.e., tweets with the tag RT. Example tweets taken from our dataset are shown in Figure \ref{fig:sampleuncleantweets}. As can be seen in the figure, the uncleaned tweets contain unwanted elements such as symbols, emoticons, numbers, punctuation marks, etc. These and other unwanted elements were removed from the data set as previously described. Below we present EDA results for the uncleaned and/or cleaned tweets.

\begin{figure}[h!]
	\centering
	\includegraphics[width=0.5\textwidth]{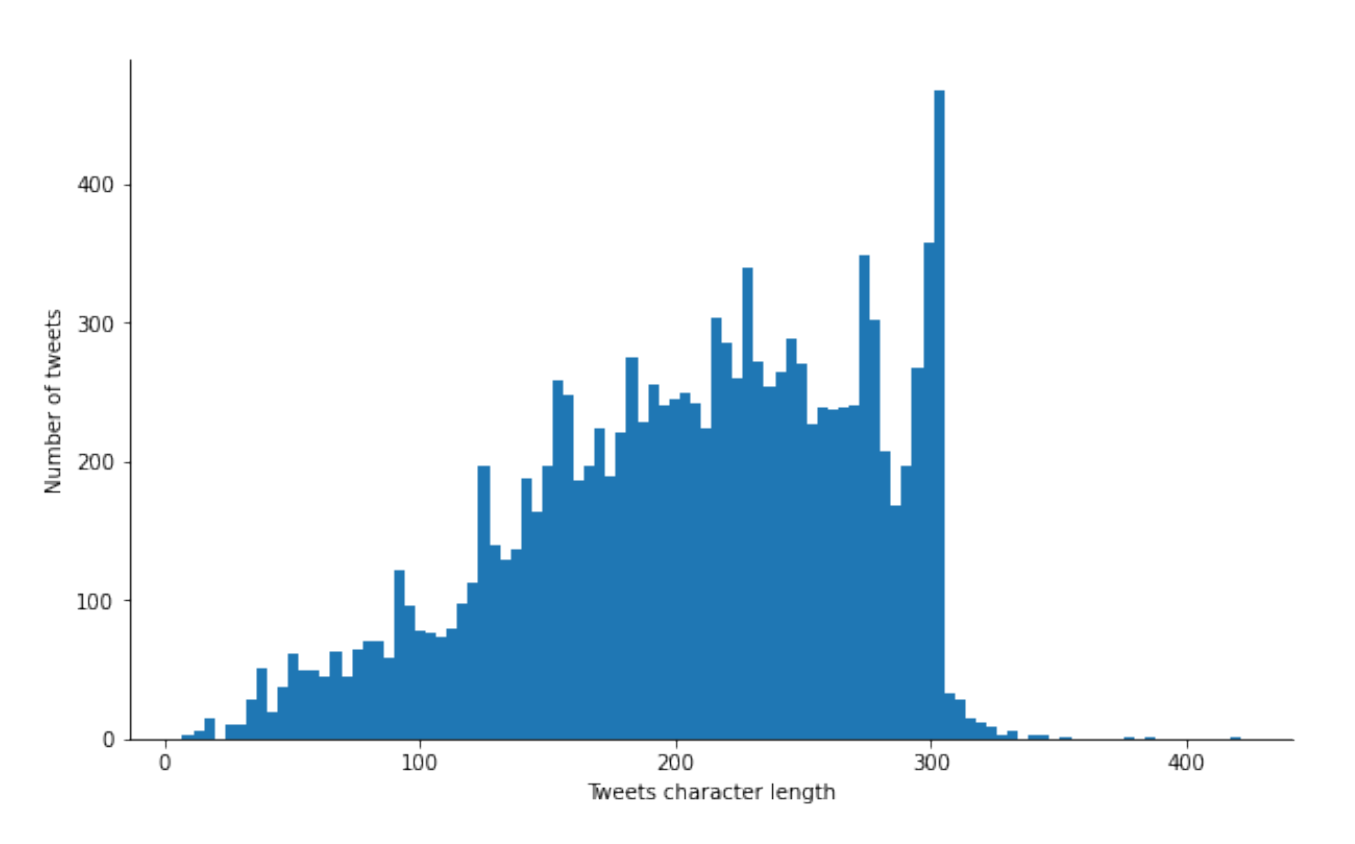}
	\caption{Tweet character length plot.}
	\label{fig:tweetcharacterlength}
\end{figure}

Figure \ref{fig:tweetcharacterlength} shows histogram plot for tweet character length. The results show that tweet character length is skewed to the right. What this means is that a large proportion of tweets tended to exploit the maximum allowable character length of 280.

\begin{figure}[h!]
	\centering
	\includegraphics[width=0.5\textwidth]{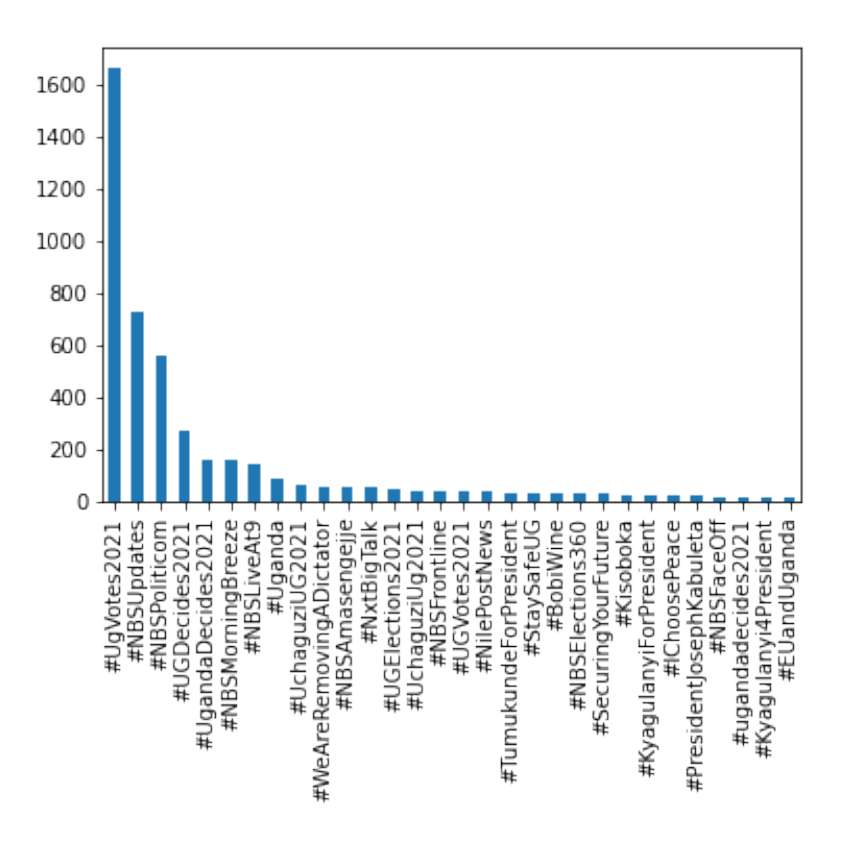}
	\caption{Frequency plot for top 30 hashtags that trended on Twitter during Uganda's 2020/2021 general elections.}
	\label{fig:hashtagfrequency}
\end{figure}

A frequency plot for hashtags in our data set is shown in Figure \ref{fig:hashtagfrequency}. The top three hashtags in this dataset include \#UgVotes2021, \#NBSUpdates, and \#NBSPoliticom. However, at least three hashtags associated with matters of public interest also trended in our dataset, for example \#WearAMask, \#StaySafeUG, and \#COVID19.

\begin{figure}[h!]
	\centering
	\includegraphics[width=0.5\textwidth]{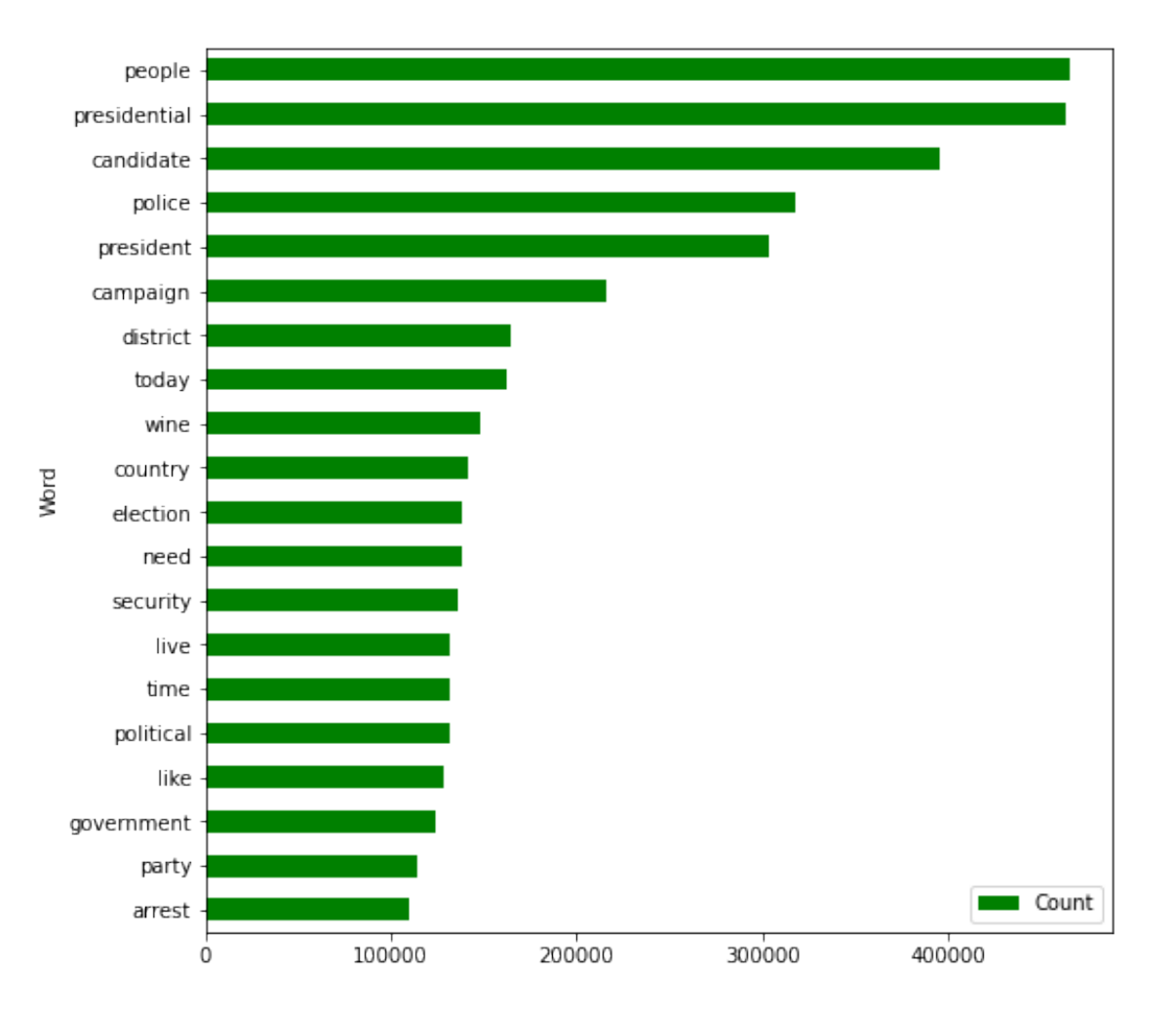}
	\caption{The top 20 frequently used words in our Twitter dataset for Uganda's 2020/2021 general elections.}
	\label{fig:frequentwords}
\end{figure}

Figure \ref{fig:frequentwords} shows the bar plot for top 20 frequently used terms/words in our Twitter dataset. The results show that the six dominant words were people, presidential, candidate, police, president, and campaign. Indeed, these words are consistent with the general context of this study as described in sub-section \ref{studycontext}.

\begin{figure}[h!]
	\centering
	\includegraphics[width=0.4\textwidth]{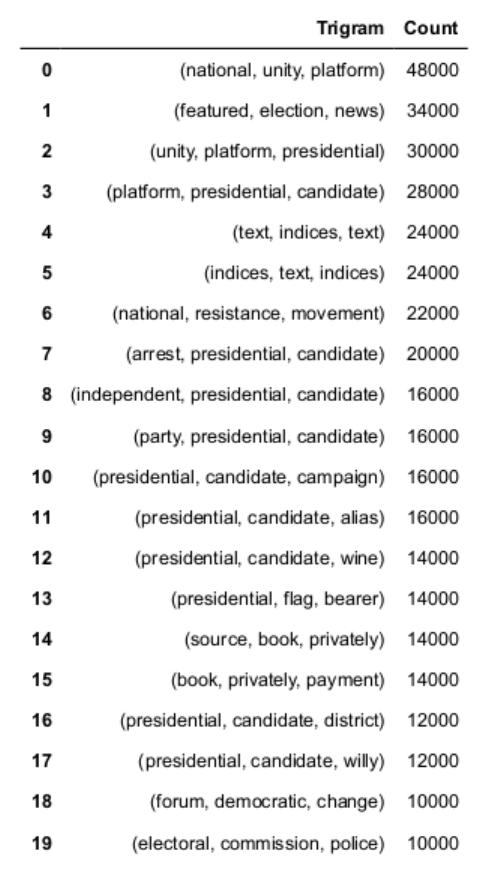}
	\caption{Top 20 frequently occurring trigrams in our Twitter data set for Uganda's 2020/2021 general elections.}
	\label{fig:trigrams}
\end{figure}

Words that often occurred together in groups of three (i.e., trigrams) in our Twitter dataset are shown in Figure \ref{fig:trigrams}. As can be seen, some of the trigrams may be associated with Uganda's 2020/2021 general elections. Examples include name of a political party (e.g., national unity platform, national resistance movement, etc) that fielded a party presidential candidate in those elections,  or events that took place during those elections (e.g., arrest of presidential candidates).

\subsection{Keywords and public policy issues}

\subsubsection{LDA topics and keywords}

\begin{figure}[h!]
	\centering
	\includegraphics[width=0.45\textwidth]{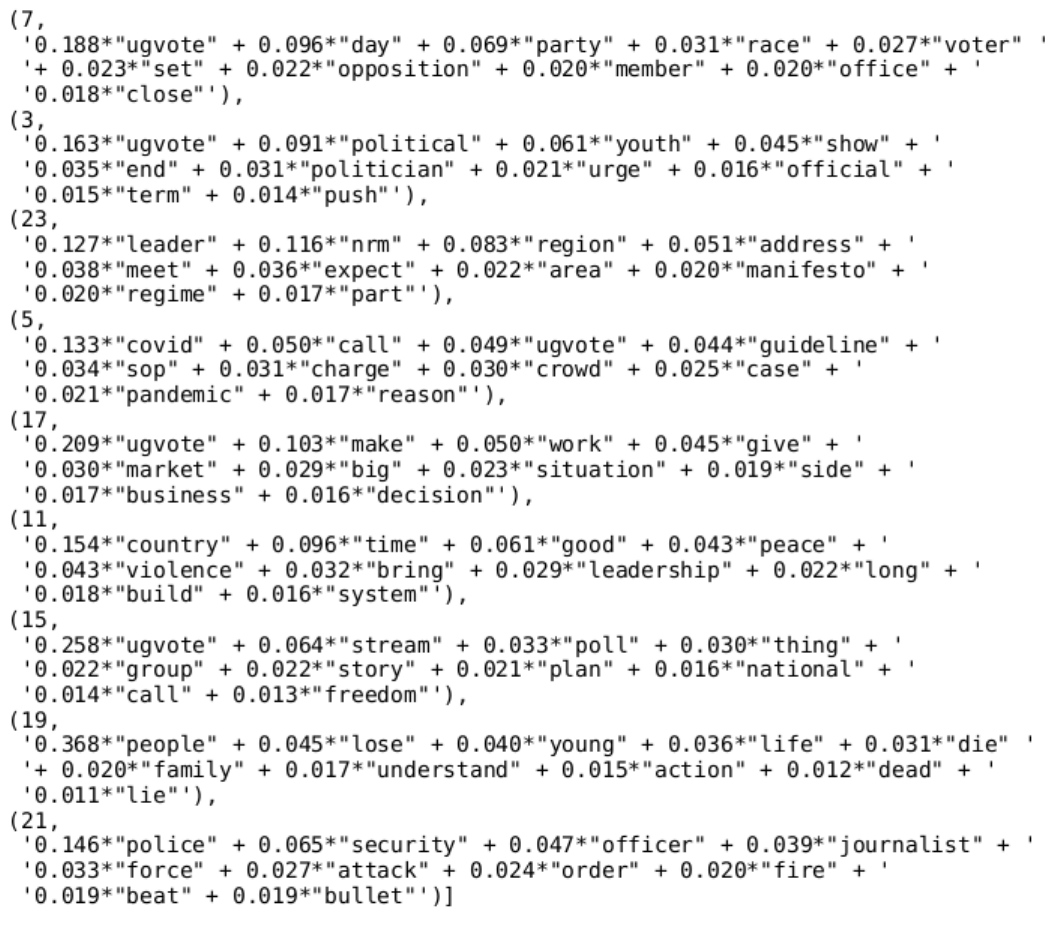}
	\caption{Some topics identified by the LDA method with their corresponding keywords.}
	\label{fig:topics}
\end{figure}

An extract of topics identified by the LDA method and their corresponding keywords is shown in Figure \ref{fig:topics}. Keywords in some of the topics are indeed semantically associated with matters of public interest, especially in context of Uganda's 2020/2021 general elections. Take for example topic 11 with keywords like peace, violence, leadership, and topic 21 with police, security, journalist, beat, fire, bullet. These keywords are descriptive of and consistent with some of the events that took place during those elections.

\begin{figure}[h!]
	\centering
	\includegraphics[width=0.45\textwidth]{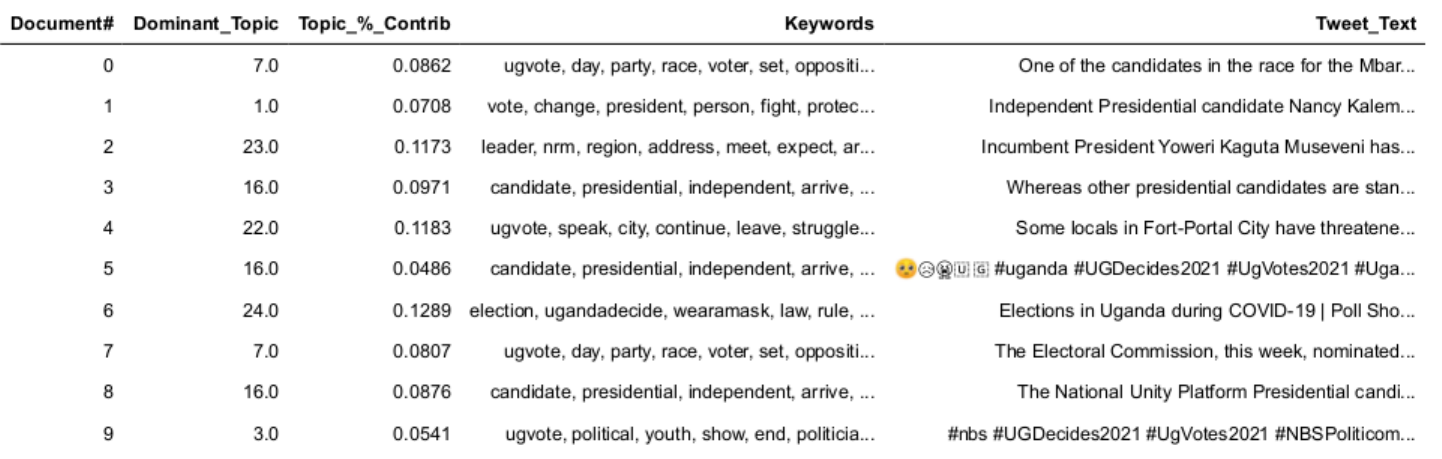}
	\caption{Dominant topic in each tweet and its percentage contribution.}
	\label{fig:dominanttopic}
\end{figure}

We determined topics that were dominant in each tweet, and computed their percentage contribution. The results are shown in Figure \ref{fig:dominanttopic}.

\begin{figure}[h!]
	\centering
	\includegraphics[width=0.45\textwidth]{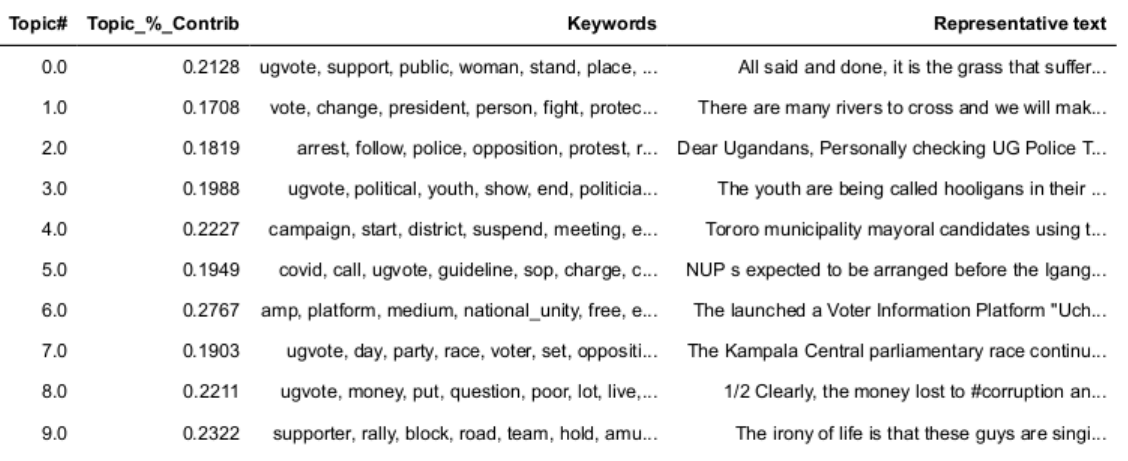}
	\caption{Most representative tweet for each LDA topic extracted from our Twitter data set.}
	\label{fig:representativetweet}
\end{figure}

Sometimes the LDA keywords are not sufficient to provide insight into what a topic is all about. An alternative way to gain understanding is to find documents (in our case, tweets) a given topic has contributed most to, and deduce the topic by reading those documents (tweets). Figure \ref{fig:representativetweet} shows the most representative tweet for each topic.

\begin{figure}[h!]
	\centering
	\includegraphics[width=0.45\textwidth]{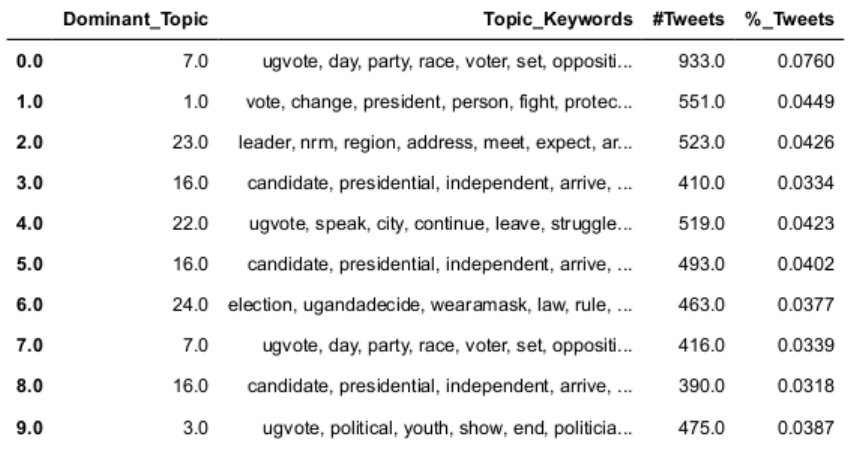}
	\caption{Topic distribution across tweets in our dataset.}
	\label{fig:topicdistribution}
\end{figure}

It is possible to determine the volume and distribution of topics to be able to judge how widely it is discussed in a dataset. The result of this analysis is shown in Figure \ref{fig:topicdistribution}.

\begin{figure}[h!]
	\centering
	\includegraphics[width=0.45\textwidth]{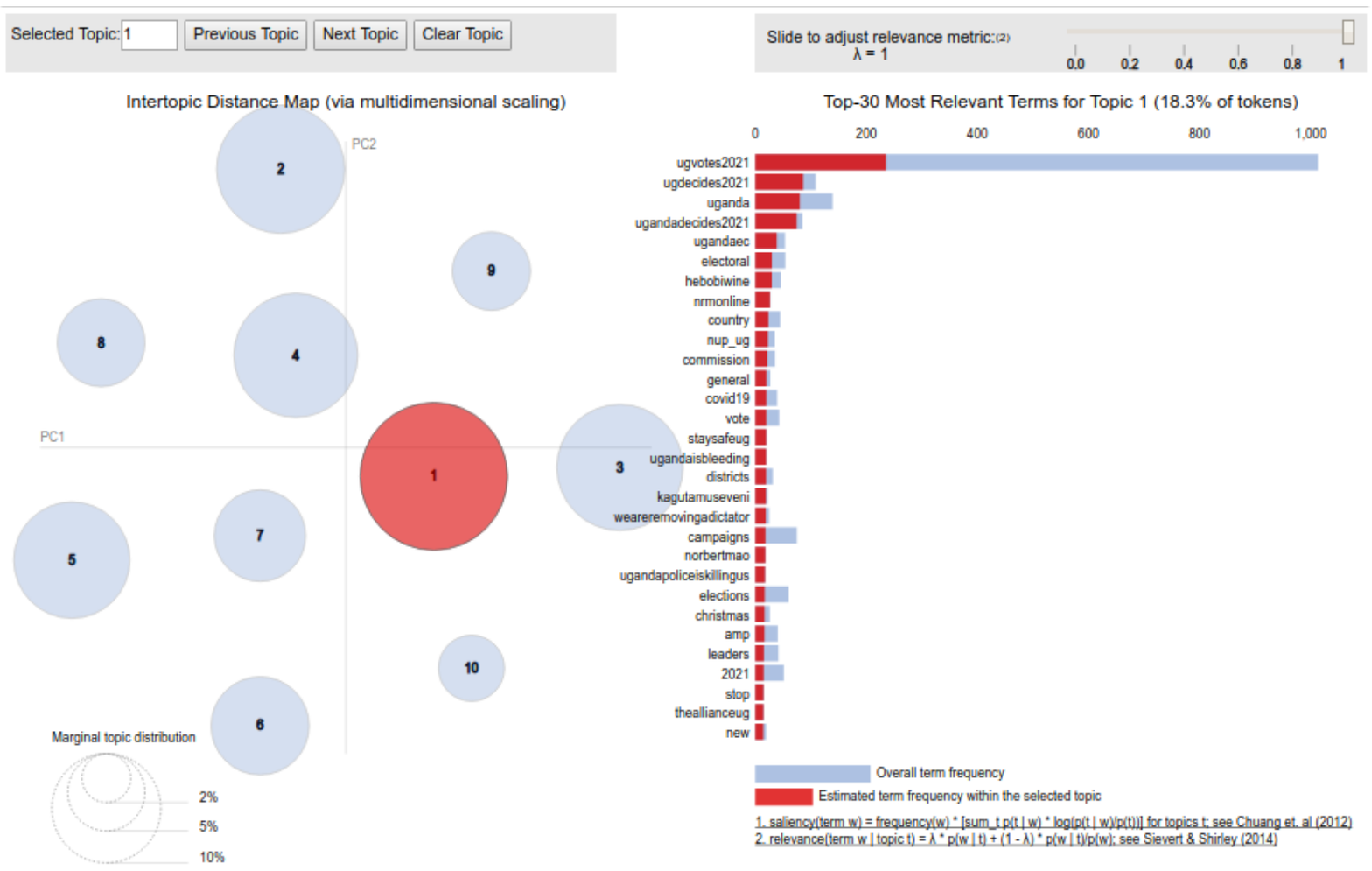}
	\caption{pyLDAvis visualization of 10 LDA topics with their corresponding top 30 keywords. The red bubble shows Topic 1 being highlighted.}
	\label{fig:topicvisualization}
\end{figure}

To visualize the identified topics along with their corresponding top $n$ keywords we used the Python library pyLDAvis \cite{Sievert2014}. Figure \ref{fig:topicvisualization} shows visualization of ten LDA topics along with their corresponding top 30 keywords. The size of the bubble is proportionate to the popularity of the topic in our dataset. These results show that there is no overlap between the identified topics.

\subsubsection{Top2Vec topics and keywords}

\begin{figure}[h]
	\centering
	\begin{subfigure}{0.4\textwidth}
		\centering
		\includegraphics[width=1.1\textwidth]{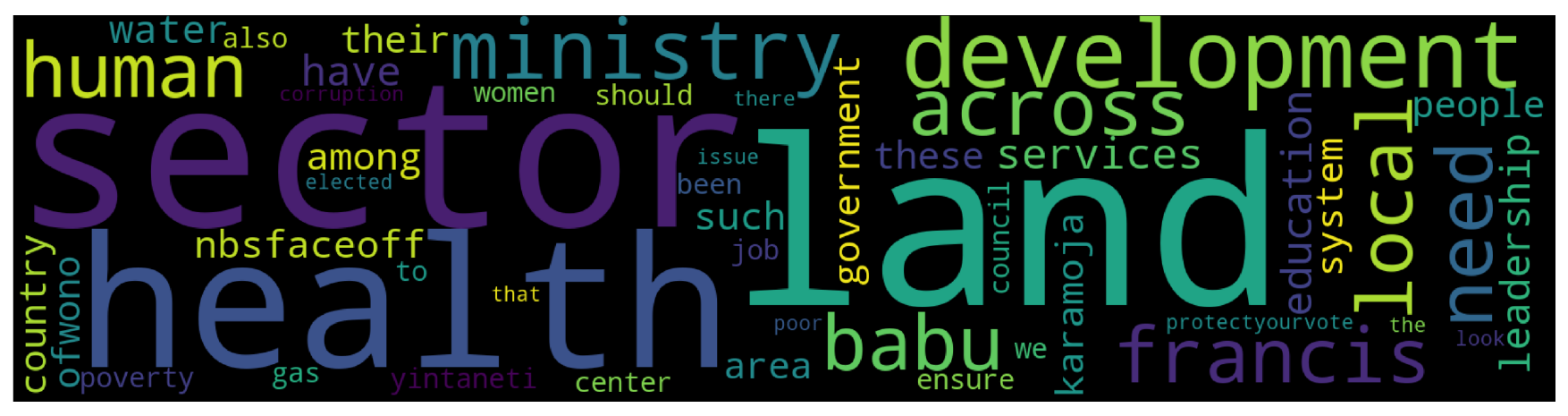}
		\caption{Topic 26.}
		\label{fig:topic26}
	\end{subfigure}
	\begin{subfigure}{0.4\textwidth}
		\centering
		\includegraphics[width=1.1\textwidth]{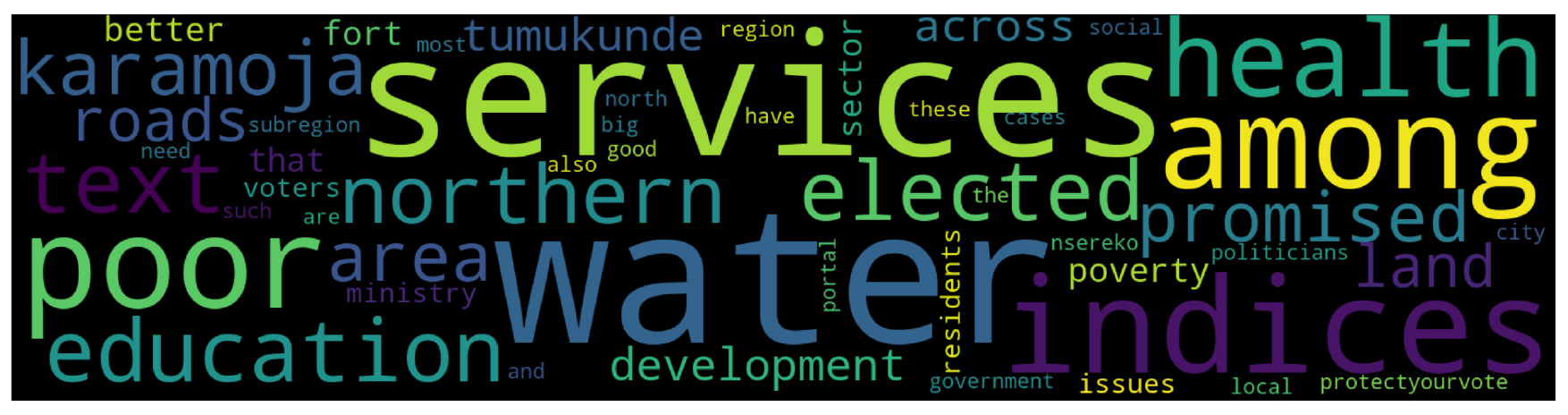}
		\caption{Topic 39.}
		\label{fig:topic39}
	\end{subfigure}
	\caption{Word clouds of top 50 keywords in Topic 26 and 39 learned using the Top2vec method. The keywords in these two topics may be semantically closely associated with the search term 'services', which was used to retrieve them from a collection of 124 identified topics.}
	\label{fig:wordclouds}
\end{figure}

The top2vec model identified a total of 124 topics from our Twitter dataset. The number of documents (i.e., tweets) associated with topics ranged between 407 (highest) and 34 (lowest). Top2vec allows the user to search for topics based on search terms. We conducted a search for the top five topics related to the term 'services'. We chose this particular phrase based on the assumption that a reasonable proportion of social media conversations during a country's general elections would focus on issues related to service delivery to citizens. Figure \ref{fig:wordclouds} shows word clouds (top 50 keywords) for the top two topics (Topic 26 and 39) associated with the services phrase. As can be seen, the keywords belonging to topics associated with this phrase are quite intuitive. Notable 10 keywords here include, water, roads, education, health, land, poverty, development, poor, leadership, and government.

\begin{figure}[h!]
	\centering
	\includegraphics[width=0.45\textwidth]{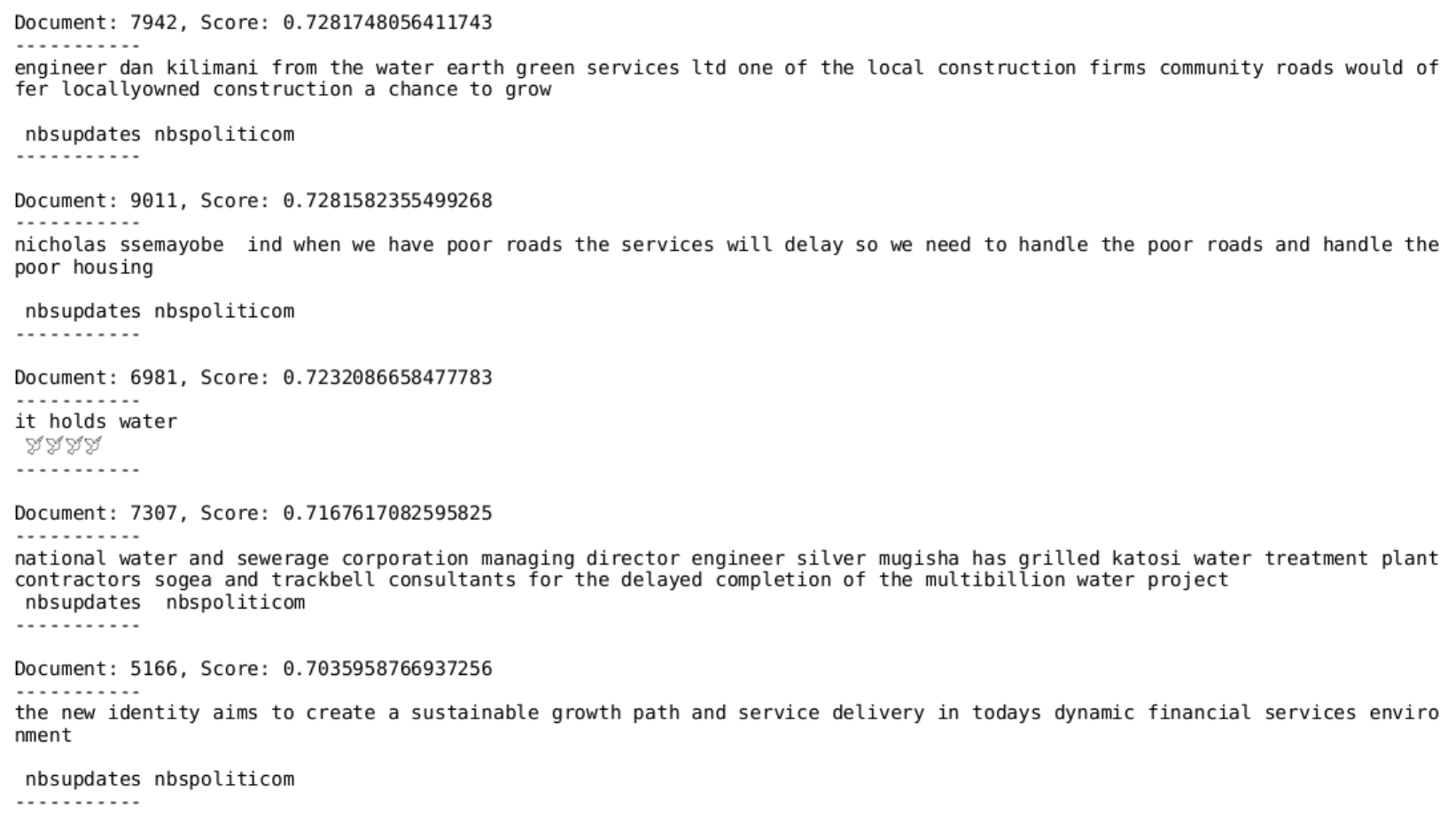}
	\caption{Document (i.e., tweet) identifier, cosine similarity score, and document text for the top five tweets associated with Topic 39, which is related most to the search term 'services'. As can be seen in each text, either the word 'service' or connotations of its meaning are apparent.}
	\label{fig:topic39-details}
\end{figure}

We retrieved the top five documents (i.e., tweets) associated with Topic 39, the top-ranked topic relating to the search term 'services'. As seen in Figure \ref{fig:topic39-details}, indeed some of the words used in the tweets are the same as those that appear in topic 39 (see Figure \ref{fig:wordclouds}). The cosine similarity of each of the five retrieved tweets associated with Topic 39 is high (at least 70\%).

Having extracted topics along with their corresponding keywords from our dataset, we proceeded to identify issue-specific keywords from them that we considered relevant for public policy attention. These issue-specific keywords were then used to create a dataset for training a GPT-2 model for text generation. The process for identifying issue-specific keywords and building a training data set is described next.

\subsubsection{Issue-specific keywords and building training data set}

Public policy related issue-specific keywords were identified from both the LDA and Top2Vec topic keywords through a manual procedure. The identification process for keywords was based on human intuition. The paragraph below presents our list of 75 issue-specific keywords:

\emph{access, army, ballot, brutality, business, change, constitution, corruption, covid19, campaigns, crowds, citizens, children, democracy, development, deaths, education, elections, fair, future, free, freedom, government, human, hospital, health, information, justice, jobs, land, law, lives, leadership, masks, media, market, military, money, opposition, pandemic, police, peace, people, protect, protests, public, politics, politicians, parties, presidential, parliament, poverty, polls, power, problems, promises, rallies, riots, rights, roads, security, services, social, situation, supporters, state, violence, vote, voter, water, war, women, work, young, youth.}

\begin{figure}[h!]
	\centering
	\includegraphics[width=0.5\textwidth]{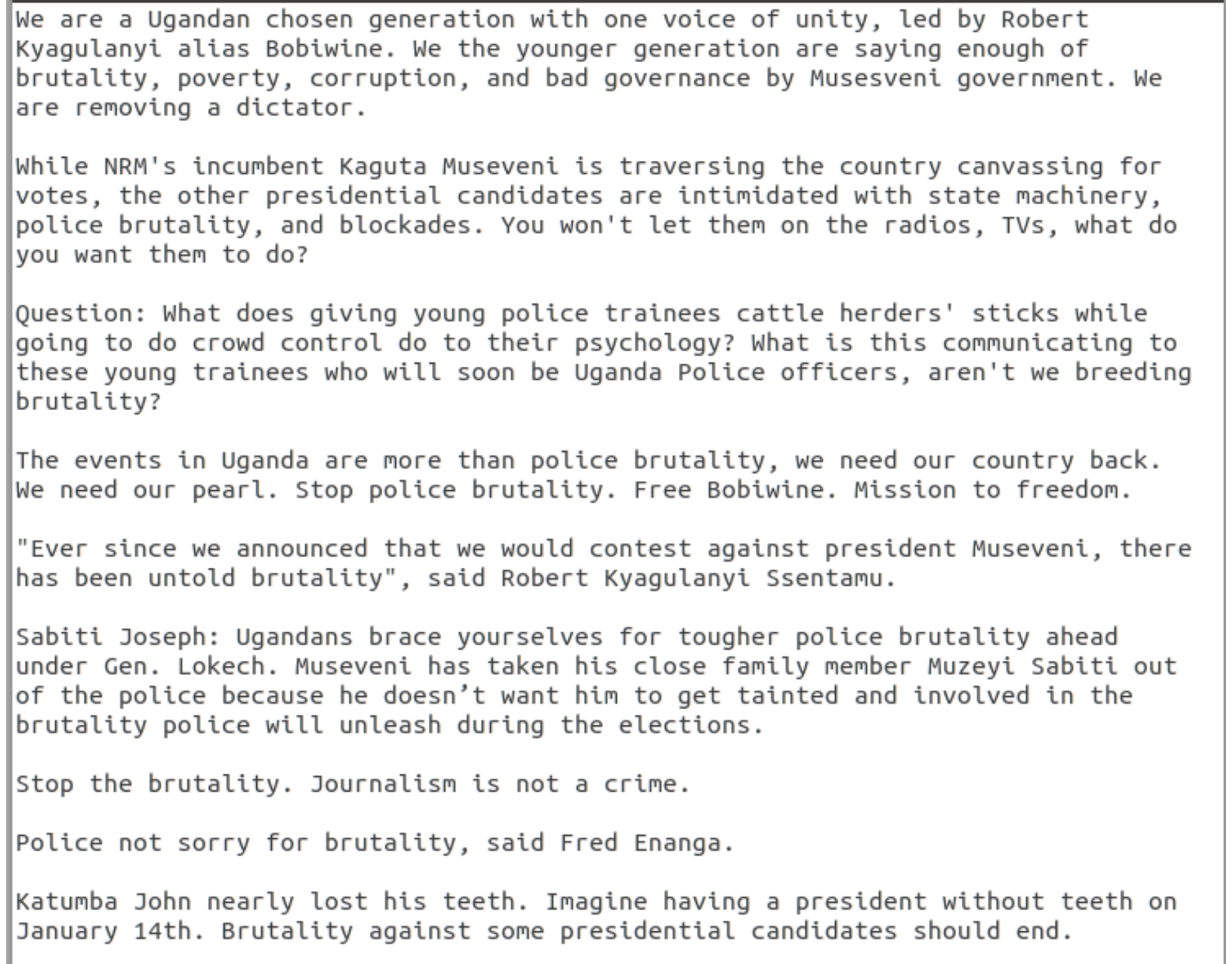}
	\caption{Example tweets associated with the theme keyword "brutality", taken from our development dataset for training a text generator model based on fine-tuning GPT-2.}
	\label{fig:tweets4brutality}
\end{figure}

We used the above keywords to retrieve relevant tweets from our text corpus. The tool we used for this task is Top2Vec's feature for semantic search of documents using keywords. For each keyword, we retrieved 100 tweets to give a total of 7,500 unprocessed tweets. These were then used to build a dataset for training a text generator model after cleaning to remove duplicate tweets, unwanted content, and correcting for grammar and punctuation. Our final model development dataset consists of 6,000 unique tweets that have semantic association with at least one theme of public interest. Figure \ref{fig:tweets4brutality} shows example tweets associated with the theme keyword "brutality".

\subsection{Public policy agenda building}

\subsubsection{Agenda text generated by our model}
\label{modelagendatext}

Sample texts generated by our model are given in Table \ref{gpt2sampletext}. Some of the output text are quite intuitive and comparable to statements made by a human. For example, in response to the prompt (prefix text) \emph{Justice, peace, equal opportunities}, the model was able to generate the phrase \emph{Justice, peace, equal opportunities, social and economic inclusion, freedom of speech, no police brutality}. However, the model also produced some seemingly incoherent pieces of text. This was seen when it was provided with the prompt \emph{Young people}, for which it output the sentence, \emph{Young people like to talk about transition and give speeches, no presidential candidate should ever be held responsible for the chaos that could follow a political debate}. Some of the texts generated by the model also seemed grammatically correct (we report results of text validation in the next sub-section).

\begin{table*}[t]
	\begin{center}
		\begin{tabular}{ll}
			{Prefix text} &{Example text generated by our GPT-2 model} \\ \hline \\
			\multirow{2}{*}{}Justice, peace, equal opportunities & Justice, peace, equal opportunities, \emph{social responsibility, and economic inclusion}. \\
			Young people &Young people \emph{want an end to police brutality and democracy promotion}. \\
			\multirow{2}{*}{}Democracy and good governance & Democracy and good governance \emph{is a human rights issue. Government should address} \\ & \emph{the issue of education so that students are employable." said Disma Nkundana}. \\
			Healthcare & Healthcare \emph{is a human right}. \\
			\multirow{2}{*}{}Road infrastructure & Road infrastructure \emph{projects across the country has been helped by the successful launch} \\ & \emph{of the country's second biometric voter education platform}. \\
		\end{tabular}
	\caption{Selected sample text generated by GPT-2 fined-tuned on our custom data set. The prefix text is the "prompt" we provided to the model while text in italics is what the model generated, given the prefix text.}~\label{gpt2sampletext}
	\end{center}
\end{table*}

\subsubsection{Validation of model text}
\label{validatedmodeltext}

35 respondents randomly selected from among 50, participated in evaluating 30 short texts generated by our GPT-2 model. Most of them (80\%) were male while 20\% female. All 35 participants held high school certificate, and were all aged between 18 and 35 years.

Results for human evaluation are shown in Table \ref{textkappas}. Based on the interpretation in \cite{Altman2006, Landis1977}, the results show "very good" inter-rater agreement on readability (K=0.78) and "good" performance on coherence (K=0.6). The Kappa coefficients were both statistically significantly different from zero.

\begin{table}[h]
	\begin{center}
		\begin{tabular}{llll}
			{Criterion}  &{Fleiss' K}
			&{95\% CI}&{$p$-value} \\ \hline \\
			
			Readability &0.78 &0.71 - 0.82 &0.000  \\
			Coherence &0.60 &0.53 - 0.63 &0.001 \\
			
		\end{tabular}
		\caption{Fleiss' Kappa statistics for human evaluation of text generated by our GPT-2 model.}~\label{textkappas}
	\end{center}
\end{table}

\subsubsection{Validation of agenda}
\label{validatedagenda}

This section presents results from validating public policy agenda items generated by our GPT-2 model. Table \ref{agendaitems} presents texts of selected agenda items formed by the model under five broad themes of economy, healthcare, democracy and good governance, education, and security. These themes were those under which the political party manifestos were summarized on the Uchaguzi website. The table also presents reference texts used in our similarity analysis. Reference text 1 and reference text 2 were extracted from the NRM and NUP manifestos, respectively.

  \begin{table*}[t]
	\centering
	\caption{GPT-2 generated agenda items and political party manifesto text extracts from Uganda's 2020/2021 general elections. Reference text 1 (NRM manifesto) and reference text 2 (NUP manifesto).}~\label{agendaitems}
	
	\small
	\begin{tabular}{*{4}{p{.20\linewidth}}}
		\toprule
		Theme & GPT-2 agenda text & Reference text 1 & Reference text 2 \\\midrule
		
		Economy & Economy and human capital teams work to ensure that development patterns and development of the country continue to improve. & Create jobs, harness production, accelerate industrialization, diversify economy, import substitution, value addition. & Inclusive economic development, equitable wealth distribution, address bottlenecks to economic development (financing, skills), infrastructure (energy, roads), and competition from low-cost producers.\\
		
		Healthcare & Healthcare is a human right. & Upgrade health centers, improve health worker welfare, improve access to specialized healthcare, promote ICT use in healthcare. & Healthcare financing policy, sustainable universal access to quality healthcare, health insurance, debt relief for health financing.\\
		
		Democracy \& good governance & Human rights, employable graduates. & Consolidate and deepen democracy, devolve power to local governments, enhance capacity of local governments to deliver services. & Restore presidential term limits, government rooted in rule of law, respect for human rights and dignity, promotion of justice and fairness, transparency and accountability of state institutions.\\
		
		Education & Education and social development minister Janet Kataha Museveni has defended the introduction of mobile telephones in schools as a way of encouraging children to benefit from the socio economic and cultural development of the country. & Promote science subjects, skills development, promote innovation and research, align higher education programs with market demands. & Transform education system, equip and staff schools, develop and promote early childhood education.\\
		
		Security & Security is our top priority. We have made sure that Ugandans have access to medicines, hospitals, clean water, and medical facilities. We can only achieve this through regular, free, and fair elections. & Train pro-people security forces, fight crime, improve welfare of security forces, continue to install CCTV cameras to fight crime. & Restore autonomy of the Uganda Police, establish independent police authority, end repression of citizens by the army, end the use of army to prop the ruling party in power.\\
		
	\end{tabular}
\end{table*}

As indicated earlier, validation of agenda items generated by our model was done through both analyzing document similarity and human evaluation. Results of the similarity analysis are shown in Table \ref{agendasimilarity}. Basing on a cosine similarity score (CSS) threshold value of 0.5 (equivalent to guess work), our model was able to generate agenda text consistent with reference text 1 on the issue of security (CSS=0.6197) and reference text 2 on both democracy and good governance (CSS=0.7633) and education (CSS=0.5075). However, the model's text prediction was below average on the issues of economy and healthcare for both reference texts.

\begin{table}[h]
	\begin{center}
		\begin{tabular}{llll}
			{Theme} &{vs. ref text 1} &{vs. ref text 2} \\ \hline \\
			
			Economy & 0.3761 & 0.2668 \\
			Healthcare & 0.3538 & 0.3820 \\
			Democracy/governance & 0.000 & {\bf 0.7633} \\
			Education & 0.3105 & {\bf 0.5075} \\
			Security & {\bf 0.6197} & 0.000 \\
			
		\end{tabular}
		\caption{Cosine similarity scores from analyzing agenda items generated by our GPT-2 model against reference texts (ref text). Score values highlighted in bold text indicate above average (0.5) similarity.}~\label{agendasimilarity}
	\end{center}
\end{table}

On the other hand, a total of 40 individuals took part in the survey to evaluate 30 agenda items generated by our GPT-2 model. Majority participants i.e., 55\% were female, 42.5\% male while 2.5\% chose not to disclose their gender identity. Participants with a masters degree were 57.5\%, bachelors 35\%, and doctorate 7.5\%. Most respondents (55\%) were aged 18 to 35 years, while the rest were 36 to 50.

Human evaluation of agenda items showed "good" agreement among raters, with K=0.450 (95\% CI 0.441 to 0.453) \cite{Altman2006, Landis1977}. The coefficient was statistically significantly different from zero ($p$=0.000).

\section{Discussion}
\label{discussion}

The goal of this work was to apply machine learning methods to the task of identifying issues from social media conversations for inclusion in public policy agenda. To achieve this goal, we proposed a 5-stage human-augmented ML pipeline that takes as input social media data that undergoes pre-processing before topic modeling is applied to extract keywords from which issues of interest are identified. The next stage involves generating agenda narratives based on the identified issue-specific keywords by employing a NLG system. The generated text is validated by human evaluators in the $4^{th}$ stage. Public interest agenda items are then derived from the narrative text and validated by human experts in the final stage. We implemented experiments to serve as proof-of-concept i.e., to validate the output of our proposed pipeline.

Results from validating text generated by our model achieved "very good" and "good" inter-rater agreement on readability and coherence, respectively. On the other hand, validation of agenda items showed reasonable results. For example, we obtained above average cosine similarity scores on at least three out of five reference texts. Human evaluation of agenda items also showed "good" agreement among raters. These results demonstrate the potential of our ML approach for identifying matters of public interest from social media conversations.

Nonetheless, our ML method did not perform as expected fully on the quality of text generated. For example some of the text produced by the model was found to be incoherent, thus achieving only "good" inter-rater agreement (see Table \ref{textkappas}). Similarity analysis also showed less than average CSS for two out of five reference text themes. Some reasons why our model may not have performed as expected include the following. Issues related to input data inconsistencies and incoherent nature of messages found in social media data sets. For example, Twitter is designed for interaction through posting short messages of up to 280 characters long as at the time of our data collection. The process for extracting tweet data off the Twitter platform involves aggregating hundreds to millions of individual tweets carrying many different conversations on diverse topics. And since the tweets are not usually organized by topic of conversation within the extracted data set, there is bound to be incoherent statements throughout the dataset. This issue is explored in \cite{Alvarez-Melis2016, Steinskog2017}. Another potential pitfall of this approach is use of unrepresentative data sets (i.e., biased data), as explained in \cite{Giest2020}. Therefore, fairness, accountability, transparency and ethical considerations are important when using such an approach as basis for policy decision-making. We also found that agenda texts generated by our NLG system had limited explanatory power i.e., carry little information as regards semantics, meaning, and context.

The pitfalls presented in the previous paragraph not withstanding, some literature from social science and qualitative policy research point out that TM may play an important role in those fields. For example \cite{Isoaho2021} contend that TM allows for application of policy theories and concepts to larger datasets. On the other hand, \cite{Paeaekkoenen2020} argue that such methods support humanistic interpretation by aiding the discovery of unexpected information in large and diverse datasets, in addition to enhancing transparency of the interpretation process.

\section{Conclusion}
\label{conclusion}

In this work we propose a ML-based human-augmented approach for generating public policy agenda from social media conversations. The approach consists of 5 stages namely, input data cleaning and pre-processing, keyword extraction and issue identification, text generation, text validation, and agenda validation. Empirical results from our work demonstrate that the ML approach represents a promising methodology for identifying matters of public interest from social media conversations. However, additional work is required to achieve higher quality generated text. Considering input data from multiple diverse social media platforms or the Web also represents an interesting new research direction. Automation of the validation stages in our pipeline could also be investigated.

\balance{}

\bibliographystyle{SIGCHI-Reference-Format}
\bibliography{sentiment}

\end{document}